\documentclass{article}
\usepackage{spconf,amsmath,graphicx}
\usepackage{amssymb}
\usepackage{algorithm}
\usepackage{algpseudocode}
\usepackage{url}
\usepackage{booktabs}
\usepackage{amsmath}
\usepackage{xcolor} 
\usepackage{enumitem}
\setlist{nosep, leftmargin=14pt}

\usepackage{mwe} 

\title{DiffKAN-Inpainting: KAN-based Diffusion model for brain tumor inpainting}
\name{\parbox{\linewidth}{\centering Tianli Tao$^{1,3,\dagger}$\thanks{$\dagger$ indicates co-first authors and * indicates corresponding author (e-mail: l.zhang.16@bham.ac.uk). The work has been partially funded by HDR UK and EU Horizon INSAFEDARE Project - Grant Agreement number: 101095661. This work is also partially supported by the STI 2030-Major Projects (No. 2022ZD0209000), National Natural Science Foundation of China (No. 62203355), Shanghai Pilot Program for Basic Research - Chinese Academy of Science, Shanghai Branch (No. JCYJ-SHFY-2022-014), Shenzhen Science and Technology Program (No. KCXFZ20211020163408012), and Shanghai Pujiang Program (No. 21PJ1421400).} \quad Ziyang Wang$^{2,\dagger}$ \quad Han Zhang$^{1}$ \quad Theodoros N. Arvanitis$^{4}$ \quad Le Zhang$^{4,*}$}}
\address{$^{1}$ School of Biomedical Engineering, ShanghaiTech University, Shanghai, China \\
    $^{2}$ The Alan Turing Institute, London, UK \\
    $^{3}$ School of Engineering, University of Birmingham, Birmingham, UK\\
    $^{4}$Digital Healthcare and Medical Imaging Research Group, School of Engineering,\\ College of Engineering and Physical Sciences, University of Birmingham, Birmingham, UK}
\begin{document}
\maketitle
\begin{abstract}
Brain tumors delay the standard preprocessing workflow for further examination. Brain inpainting offers a viable, although difficult, solution for tumor tissue processing, which is necessary to improve the precision of the diagnosis and treatment. Most conventional U-Net-based generative models, however, often face challenges in capturing the complex, nonlinear latent representations inherent in brain imaging.  In order to accomplish high-quality healthy brain tissue reconstruction, this work proposes DiffKAN-Inpainting, an innovative method that blends diffusion models with the Kolmogorov-Arnold Networks architecture. During the denoising process, we introduce the RePaint method and tumor information to generate images with a higher fidelity and smoother margin. Both qualitative and quantitative results demonstrate that as compared to the state-of-the-art methods, our proposed DiffKAN-Inpainting inpaints more detailed and realistic reconstructions on the BraTS dataset. The knowledge gained from ablation study provide insights for future research to balance performance with computing cost.
\end{abstract}
\section{Introduction}
\label{sec:intro}
Brain tumors, including gliomas, are characterized by their rapid advancement and surgical intricacy, posing considerable obstacles in treatment \cite{amin2022brain}. Magnetic resonance imaging (MRI) is the predominant method for evaluating and diagnosing diverse brain tumors, owing to its non-invasive characteristics and superior resolution for soft tissue visualization. Nonetheless, MRIs of tumor brain frequently encounter difficulties in navigating conventional public preprocessing methods employed for the quantitative analysis of brain tissue \cite{kofler2020brats}. Consequently, substituting the tumor with healthy tissue in a plausible manner is essential for subsequent research. In the field of computer vision, this approach is known as inpainting, which is why we refer to this task as brain inpainting. Furthermore, identifying the pre-lesion condition of the tumor area may enhance comprehension of the patient's tumor progression, facilitating more precise and individualized treatment approaches.

\begin{table}[t!]
\small
\centering
\caption{The parameters of different varients of U-Net.}
\begin{tabular}{c|ccc}
\toprule
Layer  & U-Net & TransUNet\cite{chen2024transunet} & \textcolor{black}{\bf U-KAN}\cite{li2024u} \\

\hline
Par$_{10^6}$& 1.08 & 87.19  & \textcolor{black}{\bf 0.33}\\
\bottomrule
\end{tabular}

\label{basicinfo}
\end{table}

Several studies have used generative adversarial network (GAN) or denoising diffusion probabilistic model (DDPM) with U-Net as their generative backbone for brain inpainting \cite{zhu2024advancing,durrer2024denoising}. However, U-Net architectures with convolutional layers commonly encounter difficulties in properly capturing nonlinear features during the denoising process. Recently, Kolmogorov-Arnold Network (KAN) \cite{liu2024kan} has emerged as formidable computational units, demonstrating exceptional performance in nonlinear feature modeling challenges. Compared to conventional U-Net and vision transformer (ViT), U-KAN has advantages to extract the nonlinear features with fewer parameters as shown in Table \ref{basicinfo} and has proven its effectiveness in several vision tasks, encompassing image and video categorization. 

\begin{figure*}[!t]
\centering
\includegraphics[width=1\linewidth]{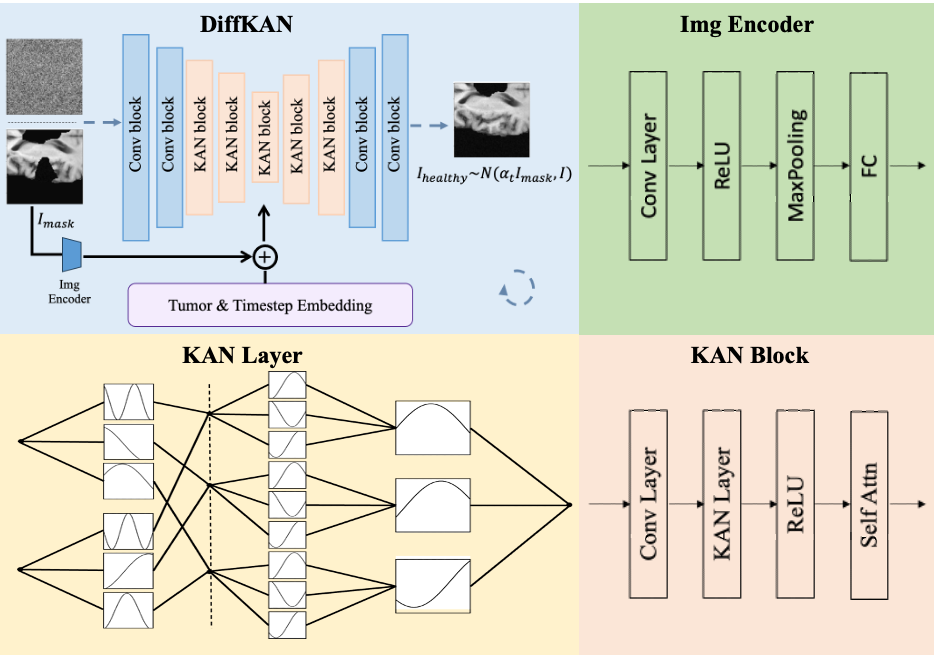} 
\caption{a) The semantic pipeline of DiffKAN. b) The structure of image encoder c) The Kolmogorov-Arnold Networks (KAN) layer. d) The structure of KAN block.}
\label{Fig:Model}
\end{figure*}

This work further explores the KAN in image inpainting task and proposed the DiffKAN-Inpainting, a U-KAN based diffusion model specifically designed for brain inpainting applications. We introduce the RePaint mechanism in the brain inpainting task, utilizing tumor position information as additional guidance for the diffusion model. DiffKAN generates personalized and anatomically plausible brain tissue that accommodates variances among participants. Additionally, we conduct an ablation study to investigate the optimal settings of the U-KAN architecture, aiming to assess the balance between computational cost and performance. These promising findings facilitate additional applications of KAN structures in medical imaging, revealing new possibilities for sophisticated picture reconstruction and analysis.

\section{Methods}

Diffusion models are generative models that learn to represnt the true distribution $p(x)$ from observations $x$, yet U-Net or Transformer based backbone remain limited by their reliance on linear modeling patterns. KAN has demonstrated their superiority and interpretability through the use of nonlinear learnable activation functions. In this work, we propose DiffKAN-Inpainting, a novel diffusion model built on the U-KAN framework for brain inpainting. Our overall pipeline is shown in Figure \ref{Fig:Model}. The latent representation of the brain is obtained using U-KAN during denoising and then inpaint the masked brain together with the tumor information.

\subsection{Overall Framework}
In this section, we will provide a complete overview of the framework. For the training stage, the DiffKAN is trained to predict the noise added to the input. This is obtained by minimizing the following objective $L$: 
\begin{equation}
L={E}\parallel\frac{X_t - X_0}{\sigma_t}-f_{\theta}(X_t, X_T, t;\theta)\parallel ,
\label{loss_func}
\end{equation}
where $f_{\theta}$ is the DiffKAN parametrized by $\theta$ and $t$ is the diffusion timestep sampled from $Uniform ({1,...,T})$. We use pairs of MRI scans denoted as $I_{healthy}$ and $I_{mask}$, obtained from the same slice with mask and without mask. For the inference stage, the model takes a masked MRI $I_{mask}$ as input and generate the inpainted image $I_{healthy}$. 

We condition the diffusion model to generate images using the following information: (i) the image representation extracted by an image encoder; (ii) the tumor geometric information. To obtain individualization at the subject level, we condition the model using a latent representation of the baseline scan. The mathematical description of the diffusion model with condition is shown in formula \ref{formula: conditional}. 
\begin{equation}
    \begin{aligned}
        p_{\theta}(x_{0:T}|c) &= p(x_{T}) \prod^{T}_{t=1} p_{\theta}(x_{t-1}|x_{t},c) ,\\
        p_{\theta}(x_{t-1}|x_{t},c) &= \mathcal{N}(x_{t-1};\mu_{\theta}(x_{t},t,c), \Sigma_{\theta}(x_{t}, t,c)),\label{formula: conditional}
    \end{aligned}
\end{equation}
where $c = p(X_{im}) + p(X_{tumor})$ is a set of paired images and text as the dual guidance. For the image guidance, the latent representation is obtained leveraging an independent encoder based on residual blocks. For the tumor information, we introduce a probability $w$ to embed the size information using a linear layer. The embedded position information is then added to the time step embedding.

\subsection{DiffKAN}
Due to the improvement of KANs, U-KAN has been developed by integrating the KAN layers into the U-shape architecture, which achieve superior performance in medical segmentation and related tasks. In this work, we implemented the U-KAN, which is mainly composed of convolution block (C) and KAN block (K). Each convolutional block is constructed by a convolution layer, a batch normalization layer and a ReLU activation layer. The convolution block is consisted of two convolutional layers and a max-pooling layer. The KAN block is consisted of a convolutional layer, a KAN layer, ReLU and self attention layer. The KAN block can be characterized as $KAN(x) = ReLU(\Phi_{k-1}\circ\Phi_{k-2}\circ...\Phi_{0}Conv(x))$, where $\Phi_{i}$ is the i-th layer in KAN network\cite{liu2024kan}. We use the skip connections to combine the encoder and decoder.

\subsection{RePaint Mechanism}
To encourage the model to generate images that accurately reflect the expected age gap between the input and the prediction, we implement a denoising method inspired by the RePaint\cite{lugmayr2022repaint} methodology to augment the model's performance in DiffKAN. Traditionally, generating the whole image and subsequently cropping it to the original dimensions would yield unsmoothed margins. To resolve this, we manipulate solely the background image and employ it as noise-free direction, instead of utilizing the noisy masked image. During each phase of the denoising process, input from the original image is used in areas that do not necessitate inpainting. This targeted procedure guarantees that only the areas requiring inpainted are modified, while the rest of the image retains fidelity to the original input. This method enhances accuracy in reconstructions and elevates visual fidelity by retaining essential information. The integration of guided denoising with the diffusion process enables DiffKAN to attain both specific refinement and overall data coherence.

\section{Experiments and results}
{\bf Dataset and Preprocessing.} We employed the public dataset from the BraTS 2023 Local Synthesis Task \cite{kofler2023brain}. This dataset comprises T1-weighted structural MRI images, associated tumor masks, and healthy brain masks. Due to the absence of corresponding healthy tissue references for tumor tissues, only data with healthy brain masks were utilized for training and testing.  The initial image resolution is $1mm \times1mm\times1mm$, with a data size of $240 \times240\times155$. To minimize computing expenses, we partitioned each subject's 3D volume into a sequence of 2D slices and cropped to $192\times192$. We exclusively utilized slices with non-zero masks for training. The training set consists of 1,257 participants, utilizing 1,241 subjects for training and 10 for validation.

{\bf Implementation Details.} All models were trained on an A100 GPU with 40GB memory. The input size was set to $192\times192$, with a batch size of 2. For the diffusion model, the number of time steps $T$ is 1000. The learning rate was set to 1e-4 for the used Adam optimizer. For the evaluation of the validation set, we use the model saved with an exponential moving average over model parameters with a rate of 0.995.

{\bf Results.} To complement the qualitative evaluations, we conducted comprehensive quantitative analyses to objectively assess the performance of DiffKAN-inpainting against the state-of-the-art baseline methods, such as, AutoEncoder \cite{doersch2016tutorial}, Pix2Pix GAN \cite{isola2017image} and DDPM \cite{ho2020denoising}. 

\begin{figure*}[h]
\centering
\includegraphics[width=1\textwidth]{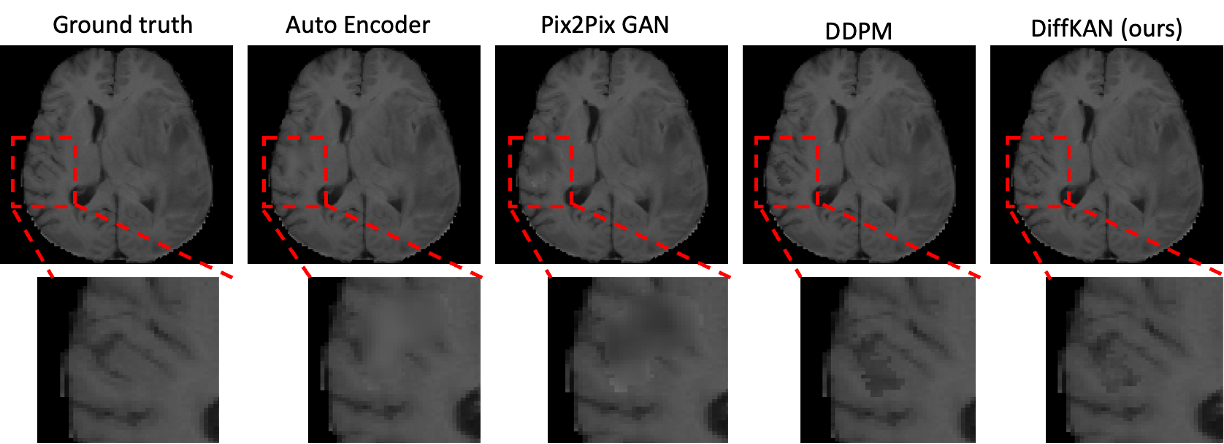} 
\caption{Showcase of the inpainted brain images by our proposed DiffKAN-Inapinting method.}
\label{Fig:result}
\end{figure*}

In Figure \ref{Fig:result}, we present a representative showcase in brain inpainting from the BraTS dataset. The results clearly demonstrate that our approach generates more realistic and intricate brain tissue structures compared to traditional U-Net-based models. The result generated by the Auto Encoder and Pix2Pix2 GAN is blurry and failed to delineate the complicated brain gyrus. Specifically, our results exhibit fewer minor disparities in the brain cerebral gyrus compared to other methods.

\begin{table}[ht] 
\centering
\setlength{\tabcolsep}{2mm}
\caption{Our proposed method has the best quantitative result.}
\begin{tabular}{ccccccccc} 
\toprule 

\multicolumn{3}{c}{}&PSNR&SSIM&MSE&\\  
\hline 
\multicolumn{3}{c}{AutoEncoder}&12.6916&0.6520&0.0934&\\   
\multicolumn{3}{c}{Pix2Pix GAN}&17.6706&0.7634&0.0288&\\
\multicolumn{3}{c}{DDPM}&17.3027&0.7416&0.0223&\\
\multicolumn{3}{c}{Ours}&\textbf{20.0588}&\textbf{0.8037}&\textbf{0.0121}&\\
\bottomrule 
\end{tabular}

\label{Table:1}
\end{table}
In Table \ref{Table:1}, we evaluate the quantitative results obtained by our proposed method and other SOTA methods using widely-adopted metrics, including peak signal-to-noise ratio (PSNR), structural similarity index (SSIM), and mean absolute error (MAE), to assess the fidelity and accuracy of the reconstructed brain tissue images. The results demonstrate that our proposed method outperforms DDPM in SSIM by +0.06, PSNR by +3 and MSE -0.01, respectively.

{\bf Ablation study.} In this section, we investigate the performance effects of different DiffKAN setups. Our objective is to achieve an balance between training time and performance. In the actual training process, we find that the training time of KAN is long, and the pure KAN network does not achieve the best performance.

\begin{table}[h] 
\centering
\setlength{\tabcolsep}{2mm}
\caption{Ablation study of different setups of DiffKAN}
\begin{tabular}{ccccccccc} 

\toprule 

\multicolumn{2}{c}{Setup}&PSNR&SSIM&MSE\\  
\hline 

\multicolumn{2}{c}{CCCCK}&19.8444&0.7938&0.0129\\
\multicolumn{2}{c}{CCCKK}&17.7634&0.7833&0.0231\\
\multicolumn{2}{c}{\textbf{CCKKK}}&\textbf{20.0588}&\textbf{0.8037}&\textbf{0.0121}\\
\multicolumn{2}{c}{CKKKK}&18.2769&0.7926&0.0174\\  
\bottomrule 
\end{tabular}

\label{Table:2}
\end{table}
In the Table \ref{Table:2}, we show the results of our method with different combinations of convolutional layer and KAN layer in the encoder.  The results have shown that the CCKKK architecture (2 Conv blocks and 3 KAN blocks in the encoder) is the optimal for the brain inapinting task. This outcome highlights that the addition of KAN layer may cause a minimal reduction in performance, indicating that the too many KAN layers contributes minimally to the performance.

\section{Conclusion}
In this work, we presented a novel KAN-based approach (DiffKAN-Inpainting) for brain tumor inpainting. RePaint mechanism and tumor information guidance were performed to simultaneously corrupt the subject-specific information to the diffusion model. Our comprehensive evaluation against state-of-the-art competing methods demonstrates DiffKAN’s superiority in terms of inpainting the brain MRI with high fidelity.

\small\bibliographystyle{IEEEbib}
\bibliography{refs}

\end{document}